\documentclass[aps, prb,
amsmath,amssymb,showpacs,groupedaddress]{revtex4}
\usepackage{graphicx}
\graphicspath{{../}}
\usepackage{epstopdf}
\begin{document}

\title{On the  flexoelectric deformations of finite  size bodies }

\author{A. S.~Yurkov}
\affiliation{644076, Omsk, Russia, e-mail: fitec@mail.ru}
\date{\today}

\pacs{77.22.-d, 77.65.-j, 77.90.+k}

\begin{abstract}

Exact equations describing flexoelectric deformation in solids,  derived  previously within the framework of a continuum media theory, are  partial differential equations of the fourth order. They are too complex to be used in the cases interesting for applications. In this paper, using the fact of smallness of the elastic moduli of a higher order,  simplified equations are proposed. Solution of the exact equations is approximately represented as a sum of two parts: the first part obeys   one-dimensional differential equations and exponentially decays  near  surface, the second ---  obeys  the equations of classical  theory of elasticity. The first part  can be constructed in an explicit  form.  For the second part, boundary conditions are obtained. They have a form of  the classical boundary conditions for the body under external forces on surface. 

\end{abstract}

\maketitle

\section{Introduction}

Although flexoelectric effect has been known for a long time (see \cite{bib:Tagantsev87,bib:Tagantsev1991} and references therein),   it  is still the subject of  intensive studies (see \cite{bib:Zubko2013,bib:Yudin2013} and references therein). In the practical sense the importance of flexoelectricity is that it, in contrast to piezoelectricity,  is symmetrically allowed in centro-symmetric materials and therefore broadens the choice of materials which can be used for electromechanical devices. Besides, although flexoelectricity  is weak  effect at macroscale, at nanoscale  it becomes  much stronger: reduced dimensions imply large gradients. From a theoretical point of view it is interesting that in the case of the flexoelectric effect there is non-trivial relation between the direct and converse effect.  There is a certain asymmetry in the bulk constitutive electromechanical equations: while  linearly varying the strain obviously contributes to the homogeneous part of the polarization, the homogeneous polarization does not appear in the equations that define the elastic stresses. Based on this, some authors have concluded that the flexoelectric effect, at least for a special arrangement, has no reversibility (see references in \cite{bib:Tag-Yur2011}). But the  general principles of thermodynamics require the effect to be reversible. This paradox is resolved as follows \cite{bib:Tag-Yur2011,bib:Yurkov2011}. The equations of elastic equilibrium are differential, so they should be appended  by  boundary conditions. In the presence of the flexoelectric effect, the elastic boundary conditions have non-classical form, and they include not only the  gradients of polarization, but also  the polarization itself \cite {bib:Yurkov2011}.  This fact leads to distortion of real body of a finite size by homogeneous polarization.

However, it is worth mentioning  that in the presence of flexoelectricity not only  the elastic boundary conditions should be changed, but also  the bulk differential equations of elasticity equilibrium should be modified \cite{bib:Yurkov2011}. The exact (within the framework of a continuum media theory) equations  are  partial differential equations of the fourth order, and they are too complex to be used in the cases interesting for applications. The only case when an analytical solution is passable is a ball of an isotropic dielectric \cite{bib:Yurkov-ball}. Even for such a simple geometry the solution is  cumbersome and requires an introduction of non-standard special functions. This is why the development of  methods of approximate description of flexoelectric deformations  is interesting. To develop one of such methods is a subject of the  present short note.

\section{Decomposition of  elastic displacements to  classical and non-classical parts and explicit  representation of non-classical part}

In  \cite{bib:Yurkov2011} equations of elastic  equilibrium and the corresponding boundary conditions were derived using the Cartesian coordinate system. However, for certain  problems the curvilinear coordinates are more appropriate. Equations in curvilinear coordinates were derived in \cite{bib:Yurkov-ball}. In such coordinates the differential equations of elastic  equilibrium are
\begin{equation}
\label{eq:eqvil0}
c^{\alpha\beta\gamma\delta}u_{\gamma;\delta;\beta} +
f^{\gamma\delta\alpha\beta } P_{\gamma;\delta;\beta } -
 v^{\alpha\beta\gamma\delta\varepsilon\zeta}u_{\gamma;\delta;\zeta;\varepsilon;\beta} = 0 \, ,
\end{equation}
where  the lower Greek indices  separated by semicolons denote covariant derivatives.  The usual partial derivatives will be denoted by lower Greek indices  separated by comma. For other notations see  \cite{bib:Yurkov-ball}.

The boundary conditions for these equations in the coordinate system where the equation of the body surface is $x^3=x^3_S={\rm const}\,$ may have different equivalent forms. Here we  assume the following form
\begin{equation}
\label{eq:bond1-0}
\Theta^{\alpha 3 3 } = 0 \, ,
\end{equation}
\begin{equation}
\label{eq:bond2-0}
\sigma^{\alpha 3} - \Theta^{\alpha \beta 3}_{\phantom{\alpha \beta 3}  ; \beta} + 
\Theta^{\alpha 3 3}_{\phantom{\alpha 3 3} , 3} +
\Theta^{\alpha (\beta\gamma)}\Gamma^3_{(\beta\gamma)} = 0 \, ,
\end{equation}
where
\begin{equation}
\sigma^{\alpha\beta} = c^{\alpha\beta\gamma\delta}u_{\gamma;\delta} +
f^{\gamma\delta\alpha\beta } P_{\gamma;\delta } -
 v^{\alpha\beta\gamma\delta\varepsilon\zeta}u_{\gamma;\delta;\zeta;\varepsilon} \,  ,
\end{equation}
\begin{equation}
\Theta^{\alpha\beta\gamma}=
v^{\alpha\beta\varepsilon\delta\gamma\zeta}u_{\varepsilon;\delta;\zeta} -
\frac{1}{2}f^{\delta\gamma\alpha\beta } P_{\delta} \, .
\end{equation}
The indices enclosed in parentheses   runs only the values 1 and 2,  if there are multiple indexes  in one pair of  parentheses then these indices are not equal to 3  simultaneously. Actually the equations (\ref{eq:bond1-0}) are equations (11), and (\ref{eq:bond2-0}) are slightly transformed equations (12) from \cite{bib:Yurkov-ball}.

When solving the problem for a ball \cite{bib:Yurkov-ball} we have seen that the solution is a sum of two parts. The first part is a  solution of  the equations of the classical theory of elasticity, the second is a non-classical part  which for small $v^{\alpha\beta\gamma\delta\varepsilon\zeta} \, $ is concentrated near the surface and decays exponentially inside the body.  Hereafter  we call the first one  a volume, or a classical part, and the second one a non-classical, or a surface part. It is naturally to assume that this property  should be preserved in  more general cases. Our aim is to use this observation for  construction of approximation that is useful for solution of more general problems than the case of a ball.

According to the observation mentioned above   we present the solution of the  equation (\ref{eq:eqvil0})  in the form  $u_{\gamma}=\tilde{u}_{\gamma}+\hat{u}_{\gamma}\,$,  where $\tilde{u}_{\gamma}$ obey the classical equations:
\begin{equation}
\label{eq:tildeeq}
c^{\alpha\beta\gamma\delta}\tilde{u}_{\gamma;\delta;\beta} +
f^{\gamma\delta\alpha\beta } P_{\gamma;\delta;\beta } = 0 \, .
\end{equation}
Using (\ref{eq:tildeeq}) and (\ref{eq:eqvil0}) we get that $\hat{u}_{\gamma}\,$ obey differential equations:
\begin{equation}
\label{eq:hateq}
c^{\alpha\beta\gamma\delta}\hat{u}_{\gamma;\delta;\beta}  -
v^{\alpha\beta\gamma\delta\varepsilon\zeta}\hat{u}_{\gamma;\delta;\zeta;\varepsilon;\beta} =
v^{\alpha\beta\gamma\delta\varepsilon\zeta}\tilde{u}_{\gamma;\delta;\zeta;\varepsilon;\beta} 
   \,  .
\end{equation}
Substitution  $u_{\gamma}=\tilde{u}_{\gamma}+\hat{u}_{\gamma}\,$ and some transformations also  yields that the boundary conditions (\ref{eq:bond1-0})  and (\ref{eq:bond2-0}) takes the form:
\begin{equation}
\label{eq:bond1m}
v^{\alpha 3\varepsilon\delta 3 \zeta}\hat{u}_{\varepsilon;\delta;\zeta} =
\frac{1}{2}f^{\delta 3 \alpha 3} P_{\delta} - 
v^{\alpha 3\varepsilon\delta 3 \zeta}\tilde{u}_{\varepsilon;\delta;\zeta} \, ,
\end{equation}
\vspace{0.3cm}
\begin{equation}
\label{eq:bond2m}
\begin{array}{l} 
\displaystyle
c^{\alpha 3\gamma\delta}\tilde{u}_{\gamma;\delta} +
f^{ \gamma\delta \alpha 3  } P_{\gamma;\delta } -
 v^{\alpha 3\gamma\delta\varepsilon\zeta}\tilde{u}_{\gamma;\delta;\zeta;\varepsilon} - 
 v^{\alpha (\beta)\varepsilon\delta 3 \zeta}\tilde{u}_{\varepsilon;\delta;\zeta ; (\beta)} +
\frac{1}{2}f^{ \delta 3 \alpha (\beta) } P_{\delta ; (\beta)}  + 
v^{\alpha(\beta)\varepsilon\delta(\gamma)\zeta}\tilde{u}_{\varepsilon;\delta;\zeta}
\Gamma^3_{(\beta)(\gamma)} - \\ \\
\displaystyle
\frac{1}{2}f^{\delta (\gamma) \alpha(\beta)  } P_{\delta} \Gamma^3_{(\beta)(\gamma)} + 
c^{\alpha 3\gamma\delta}\hat{u}_{\gamma;\delta} -
 v^{\alpha 3\gamma\delta\varepsilon\zeta}\hat{u}_{\gamma;\delta;\zeta;\varepsilon} -
  v^{\alpha (\beta)\varepsilon\delta 3 \zeta}\hat{u}_{\varepsilon;\delta;\zeta ; (\beta)} + 
v^{\alpha(\beta)\varepsilon\delta(\gamma)\zeta}\hat{u}_{\varepsilon;\delta;\zeta}
\Gamma^3_{(\beta)(\gamma)}  = 0 \, .
\end{array}
\end{equation}
\vspace{0.5cm}

Naturally decomposition  $u_{\gamma}=\tilde{u}_{\gamma}+\hat{u}_{\gamma}\,$  is not unique. By requiring that $\tilde{u}_{\gamma}$ obey (\ref{eq:tildeeq}), we have restricted this  non-uniqueness, but have not eliminated it completely. It is clear that $\tilde{u}_{\gamma} $ does not contain a non-classical part, but it does not mean that $\hat{u}_{\gamma} $ does not contain a classical part.  Thus  further restriction is  required. This restriction will be done in the following way: when constructing  the explicit form of  $\hat{u}_{\gamma}\,$, we will discard the non-exponential terms.  The details will be clear from what follows.

Below we use the fact that  real values of  $v^{\alpha\beta\gamma\delta\varepsilon\zeta} \, $ are  small. It is convenient to assume that all of them are proportional to a scalar  $v \to 0 \,$. We assume also that all $c^{\alpha\beta\gamma\delta}\,$  is proportional to a scalar $c\,$.  If $v \to 0\,$ then  the right-hand side of (\ref{eq:hateq}) and the second term on the right hand side of (\ref{eq:bond1m}) can be neglected.  Indeed since $\tilde{u}_{\gamma}$ obey the classical equation (\ref{eq:tildeeq}) in the case of sufficiently smooth surface and in the absence of polarization gradients tend to infinity in the limit $v \to 0\,$ , the classical part can not have large gradients which can compensate smallness of $v^{\alpha\beta\gamma\delta\varepsilon\zeta} \, $. 
As for the non-classical part $\hat{u}_{\gamma}\,$, the situation is different: here such compensation is possible, but only if the derivatives are in   $x^3\,$. Moreover,  in a thin layer near the surface  covariant derivatives in $x^3\,$  can be replaced by the usual  derivatives,  $\Gamma$-terms give  only small corrections  here. Thus (\ref{eq:hateq}) and  (\ref{eq:bond1m}) can be approximately replaced by
\begin{equation}
\label{eq:hateqm}
c^{\alpha 3 \gamma 3}\hat{u}_{\gamma , 3 , 3}  -
v^{\alpha 3 \gamma 3 3 3}\hat{u}_{\gamma , 3 , 3 , 3 , 3 } = 0 \, ,
\end{equation}
\begin{equation}
\label{eq:bond1m2}
v^{\alpha 3\varepsilon 3 3 3}\hat{u}_{\varepsilon , 3 , 3 } =
\frac{1}{2}f^{\delta 3 \alpha 3} P_{\delta} \, .
\end{equation}

Equations (\ref{eq:hateqm}) are a system of three ordinal differential equations, the dependence on $x^1$ and $x^2$ is parametric here. Moreover,  in a thin layer near the body surface one can assume  that the coefficients  do not depends on $x^3\,$, they are approximately equal  to the surface values. Solution of such a system is easy. In a standard way one should find a fundamental basis set of solutions in the form $\hat{u}_{\gamma}=\bar{u}_{\gamma}e^{\lambda(x^3-x^3_S)}\,$. Obviously equations for amplitudes $\bar{u}_{\gamma}$  are
\begin{equation}
\label{eq:eigen}
\lambda^2 c^{\alpha 3 \gamma 3}\bar{u}_{\gamma } =
\lambda^4 v^{\alpha 3 \gamma 3 3 3}\bar{u}_{\gamma} 
 \,  .
\end{equation}

It is clear from (\ref{eq:eigen})  that fundamental solutions with $\lambda=0$ are possible.    But it is clear also that such $\lambda$ corresponds to non-exponential, classical solutions. This is why we should exclude the case of $\lambda=0\,$. By this exclusion we completely  eliminate  the non-uniqueness of  the decomposition  $u_{\gamma}=\tilde{u}_{\gamma}+\hat{u}_{\gamma}\,$ mentioned above. Thus (\ref{eq:eigen}) should be changed to
\begin{equation}
\label{eq:eigenm}
 c^{\alpha 3 \gamma 3}\bar{u}_{\gamma } =
\lambda^2 v^{\alpha 3 \gamma 3 3 3}\bar{u}_{\gamma} 
 \,  .
\end{equation}

Equation (\ref{eq:eigenm}) is a standard generalized eigenproblem for  symmetric positive definite $ 3\times 3$ matrices.  Therefore we consider $\lambda_n$ and  $\bar{u}_{\gamma}^n$ to be  known. If the body bulk corresponds to $x^3 \leq x^3_S\,$ for definiteness, then $\lambda_n$ equals the positive square root  of the  $n$-th eigenvalue.  General representation of non-classical part is
\begin{equation}
\hat{u}_{\gamma} = \sum_{n=1}^3 a_n \bar{u}^n_{\gamma}e^{\lambda_n(x^3-x^3_S)}\, ,
\end{equation}
where only the coefficients $a_n$ are  unknown. The latter can be found easily by (\ref{eq:bond1m2}) which yields a simple system of linear algebraic equations:
\begin{equation}
\label{eq:anfind}
\sum_{n=1}^3 \lambda_n^2 v^{\alpha 3\gamma 3 3  3}\bar{u}_{\gamma}^n a_n =
\frac{1}{2}f^{\delta 3 \alpha 3} P_{\delta} \, .
\end{equation}
Note that   the smallness of $v^{\alpha 3 \gamma 3 3 3}$  in  (\ref{eq:anfind})  is compensated   by $\lambda_n^2$ and using (\ref{eq:eigenm}) this system of equations   can be rewritten  in the form:
\begin{equation}
\sum_{n=1}^3  c^{\alpha 3\gamma 3}\bar{u}_{\gamma}^n a_n =
\frac{1}{2}f^{\delta 3 \alpha 3} P_{\delta} \, .
\end{equation}
Thus the non-classical part of elastic displacements is found completely in explicit form.

\section{Boundary conditions for the classical part}

We saw above that to find  completely the non-classical part of elastic displacements one needs only the boundary conditions  (\ref{eq:bond1m}).  The remaining boundary conditions (\ref{eq:bond2m}) yield  the boundary conditions for the classical equations (\ref{eq:tildeeq}) in this  way, one should only substitute known $\hat{u}_{\gamma}$ to (\ref{eq:bond2m}). But to avoid the excess of accuracy, all the terms which  tend to zero in the limit of $v \to 0$ should be omitted in the final form of these boundary conditions. Besides, further down it  will be clear that (\ref{eq:bond2m}) contains  terms which diverge in this limit. However it is worth mentioning  that these divergent terms completely cancel each other out.  So these terms should be  omitted.

According to the above,  all the terms in (\ref{eq:bond2m}) where gradients of $\tilde{u}_{\gamma}$ are convolved with $v^{\alpha\beta\gamma\delta\varepsilon\zeta}$ should be omitted for the same reasons as in the case of the equation (\ref{eq:hateq}).  Further it is convenient to introduce the notation:
\begin{equation}
\label{eq:sdef}
s^{\alpha} =
 - c^{\alpha 3\gamma\delta}\hat{u}_{\gamma;\delta} +
 v^{\alpha 3\gamma\delta\varepsilon\zeta}\hat{u}_{\gamma;\delta;\zeta;\varepsilon} +
  v^{\alpha (\beta) \gamma \delta 3 \zeta}\hat{u}_{\gamma;\delta;\zeta ; (\beta)} - 
v^{\alpha(\beta)\varepsilon\delta(\gamma)\zeta}\hat{u}_{\varepsilon;\delta;\zeta}
\Gamma^3_{(\beta)(\gamma)}   \, .
\end{equation}
With this notation the boundary conditions to the equations of classic theory of elasticity can be written as
\begin{equation}
\label{eq:bond3m2}
c^{\alpha 3\gamma\delta}\tilde{u}_{\gamma;\delta} =  s^{\alpha} - 
 f^{\gamma\delta \alpha 3  } P_{\gamma;\delta } - 
 \frac{1}{2}f^{ \delta 3 \alpha (\beta) } P_{\delta ; (\beta)} +  
\frac{1}{2}f^{\delta (\gamma) \alpha(\beta) } P_{\delta} \Gamma^3_{(\beta)(\gamma)}
  \, .
\end{equation}

Now all the terms to be further simplified, are contained in $s^{\alpha}\,$. Carrying out  simplification,  first of all one should leave in the last term only the partial (not  covariant) derivatives in $x^3\,$. The reason for this is the same as above. Next, one should  express the remaining covariant derivatives in terms of  partial derivatives. The exact expression for the third covariant derivative of the vector is very cumbersome. But keeping in mind that these third covariant derivatives are convolved with $v^{\alpha\beta\gamma\delta\varepsilon\zeta}\,$, one can  omit all the terms  where  initial vector is not differentiated at least  two times. So   it turns out:
\begin{equation}
A_{\alpha ; \beta ; \gamma ; \zeta} \approx A_{\alpha , \beta , \gamma , \zeta}
-  A_{ \delta , \beta , \zeta}\Gamma^{\delta}_{\alpha\gamma} 
-  A_{ \delta , \gamma , \zeta}\Gamma^{\delta}_{\alpha \beta}
-  A_{ \delta , \beta , \gamma}\Gamma^{\delta}_{\alpha\zeta}
-  A_{\alpha , \delta , \zeta}\Gamma^{\delta}_{\beta\gamma}
-  A_{\alpha , \delta , \gamma}\Gamma^{\delta}_{\beta\zeta}
-  A_{\alpha , \beta ,\delta }\Gamma^{\delta}_{\gamma \zeta}
\, .
\end{equation}
With this equation and simplifications  mentioned above it turns
\begin{equation}
\label{eq:sdefm2}
\begin{array}{l}
\displaystyle
s^{\alpha} = \\ \\
\displaystyle
 - c^{\alpha 3\gamma\delta}\hat{u}_{\gamma;\delta} 
+ v^{\alpha 3\gamma\delta\varepsilon\zeta} \hat{u}_{\gamma , \delta , \zeta , \varepsilon} 
-   v^{\alpha 3\gamma\delta\varepsilon\zeta} \hat{u}_{ \rho , \delta , \varepsilon}\Gamma^{\rho}_{\gamma\zeta} 
- v^{\alpha 3\gamma\delta\varepsilon\zeta} \hat{u}_{ \rho , \zeta , \varepsilon}
\Gamma^{\rho}_{\gamma \delta} 
- v^{\alpha 3\gamma\delta\varepsilon\zeta}\hat{u}_{ \rho , \delta , \zeta}
\Gamma^{\rho}_{\gamma \varepsilon}
- v^{\alpha 3\gamma\delta\varepsilon\zeta} \hat{u}_{\gamma , \rho , \varepsilon}
\Gamma^{\rho}_{\delta \zeta} - \\ \\
\displaystyle
 v^{\alpha 3\gamma\delta\varepsilon\zeta} \hat{u}_{\gamma , \rho ,\zeta}
\Gamma^{\rho}_{\delta\varepsilon}
- v^{\alpha 3\gamma\delta\varepsilon\zeta} \hat{u}_{\gamma , \delta ,\rho }
\Gamma^{\rho}_{\zeta \varepsilon} 
+ v^{\alpha (\beta) \gamma \delta 3 \zeta}\hat{u}_{\gamma ,\delta ,\zeta , (\beta)}
 - v^{\alpha (\beta) \gamma \delta 3 \zeta}\hat{u}_{ \rho , \delta , (\beta)}
 \Gamma^{\rho}_{\gamma\zeta} 
- v^{\alpha (\beta) \gamma \delta 3 \zeta} \hat{u}_{ \rho , \zeta , (\beta)}
\Gamma^{\rho}_{\gamma \delta}  - \\ \\
\displaystyle
 v^{\alpha (\beta) \gamma \delta 3 \zeta} \hat{u}_{ \rho , \delta , \zeta}
\Gamma^{\rho}_{\gamma (\beta)}
- v^{\alpha (\beta) \gamma \delta 3 \zeta} \hat{u}_{\gamma , \rho , (\beta)}
\Gamma^{\rho}_{\delta \zeta} 
- v^{\alpha (\beta) \gamma \delta 3 \zeta} \hat{u}_{\gamma , \rho , \zeta}
\Gamma^{\rho}_{\delta (\beta)} 
-  v^{\alpha (\beta) \gamma \delta 3 \zeta} \hat{u}_{\gamma , \delta ,\rho }
\Gamma^{\rho}_{\zeta (\beta)} - \\ \\
\displaystyle
 v^{\alpha(\beta)\varepsilon 3 (\gamma) 3}\hat{u}_{\varepsilon ,3 , 3}
\Gamma^3_{(\beta)(\gamma)}   \, .
\end{array}
\end{equation}
Again, using the fact that to compensate  smallness of  $v^{\alpha\beta\gamma\delta\varepsilon\zeta}\,$ one needs  at least  two  derivatives in $x^3\,$, we further simplify this equation:
\begin{equation}
\label{eq:sdefm3}
\begin{array}{l}
\displaystyle
s^{\alpha} = \\ \\
\displaystyle
 - c^{\alpha 3\gamma\delta}\hat{u}_{\gamma;\delta} 
+ v^{\alpha 3\gamma\delta\varepsilon\zeta} \hat{u}_{\gamma , \delta , \zeta , \varepsilon} 
-   v^{\alpha 3\gamma 3 3 \zeta} \hat{u}_{ \rho ,  3 , 3 }\Gamma^{\rho}_{\gamma\zeta} 
- v^{\alpha 3\gamma\delta  3 3 } \hat{u}_{ \rho ,  3  , 3 }\Gamma^{\rho}_{\gamma \delta} 
- v^{\alpha 3\gamma 3 \varepsilon 3}\hat{u}_{ \rho , 3  , 3}\Gamma^{\rho}_{\gamma \varepsilon}
- v^{\alpha 3\gamma\delta 3\zeta} \hat{u}_{\gamma ,  3  , 3 }\Gamma^{3}_{\delta \zeta} - \\ \\
\displaystyle
 v^{\alpha 3\gamma\delta\varepsilon 3} \hat{u}_{\gamma , 3  , 3}\Gamma^{3}_{\delta\varepsilon}
- v^{\alpha 3\gamma 3 \varepsilon\zeta} \hat{u}_{\gamma ,  3 , 3 }\Gamma^{3}_{\zeta \varepsilon} 
+ v^{\alpha (\beta) \gamma 3 3 3}\hat{u}_{\gamma , 3 , 3 , (\beta)}
-  v^{\alpha (\beta) \gamma  3 3  3} \hat{u}_{ \rho ,  3 , 3}\Gamma^{\rho}_{\gamma (\beta)}
- v^{\alpha (\beta) \gamma \delta 3 3} \hat{u}_{\gamma , 3 , 3}\Gamma^{3}_{\delta (\beta)} - \\ \\
\displaystyle
v^{\alpha (\beta) \gamma 3 3 \zeta} \hat{u}_{\gamma , 3  , 3 }\Gamma^{3}_{\zeta (\beta)}
- v^{\alpha(\beta)\varepsilon 3 (\gamma) 3}\hat{u}_{\varepsilon ,3 , 3}
\Gamma^3_{(\beta)(\gamma)}   \, .
\end{array}
\end{equation}
It is convenient to introduce yet another  notation:
\begin{equation}
\label{eq:hdef}
\begin{array}{l}
\displaystyle
h^{\alpha\rho}
= v^{\alpha 3\gamma 3 3 \zeta} \Gamma^{\rho}_{\gamma\zeta} 
+ v^{\alpha 3\gamma\delta  3 3 }\Gamma^{\rho}_{\gamma \delta} 
+ v^{\alpha 3\gamma 3 \varepsilon 3}\Gamma^{\rho}_{\gamma \varepsilon}
+ v^{\alpha 3\rho\delta 3\zeta} \Gamma^{3}_{\delta \zeta} 
+ v^{\alpha 3\rho\delta\varepsilon 3}\Gamma^{3}_{\delta\varepsilon}
+ v^{\alpha 3\rho 3 \varepsilon\zeta}\Gamma^{3}_{\zeta \varepsilon} 
+  v^{\alpha (\beta) \gamma  3 3  3} \Gamma^{\rho}_{\gamma (\beta)} + \\ \\
\displaystyle
v^{\alpha (\beta) \rho \delta 3 3} \Gamma^{3}_{\delta (\beta)}
+  v^{\alpha (\beta) \rho 3 3 \zeta} \Gamma^{3}_{\zeta (\beta)}
+ v^{\alpha(\beta)\rho 3 (\gamma) 3} \Gamma^3_{(\beta)(\gamma)}   \, .
\end{array}
\end{equation}
Using the symmetry properties of $ v^{\alpha \beta \gamma \delta \varepsilon \zeta} \, $, it can be rewritten as:
\begin{equation}
\label{eq:hdefmm}
\displaystyle
h^{\alpha\beta} =
 ( v^{\alpha 3\gamma 3 \delta 3}
+ 2v^{\alpha 3\gamma\delta  3 3 })\Gamma^{\beta}_{\gamma \delta} 
+ (2v^{\alpha 3\beta \gamma\delta 3}
+ v^{\alpha 3\beta\gamma 3\delta} )\Gamma^{3}_{\gamma \delta}  
+  2v^{\alpha 3 \beta \gamma  (\delta) 3}\Gamma^{3}_{\gamma (\delta)}  
+  v^{\alpha  3 \gamma  3 (\delta)  3} \Gamma^{\beta}_{\gamma (\delta)}
+ v^{\alpha (\gamma) \beta 3 (\delta) 3 } \Gamma^3_{(\gamma)(\delta)}   \, .
\end{equation}
In this notation the expression for $s^{\alpha}$  can be written as follows:
\begin{equation}
\label{eq:sdefm4}
s^{\alpha} =  - c^{\alpha 3\gamma\delta}\hat{u}_{\gamma , \delta} + 
c^{\alpha 3\gamma\delta}\hat{u}_{\varepsilon}
\Gamma^{\varepsilon}_{\gamma\delta}
+ v^{\alpha 3\gamma\delta\varepsilon\zeta} \hat{u}_{\gamma , \delta , \zeta , \varepsilon} 
+ v^{\alpha (\beta) \gamma 3 3 3}\hat{u}_{\gamma , 3 , 3 , (\beta)} 
- h^{\alpha\beta}\hat{u}_{\beta , 3 , 3}  \, .
\end{equation}

Now one can proceed  to the analysis of the divergent terms. Obviously they appear in the first and third term, when all the derivatives here are in $x^3\,$. But it is also clear that these divergent terms exactly cancel each other out because of the explicit form of $\hat{u}_{\gamma}\,$, derived above. One might think that these divergent terms should simply be excluded, but in reality the situation is more complicated. 

The fact is that    
$- c^{\alpha 3\gamma 3}\hat{u}_{\gamma , 3 , 3}  +  v^{\alpha 3\gamma 3 3 3}
\hat{u}_{\gamma , 3 , 3 , 3 ,3}$ is equal to zero only for  the main approximation for $\hat{u}_{\gamma}\,$. But the corrections to this  approximation may result in the fact that this expression becomes finite. Henceforth we denote  such corrections   as $w_{\gamma}$ keeping notation $\hat{u}_{\gamma}$  only for the main approximation. In this notation we can write:
\begin{equation}
\label{eq:sdefm5}
s^{\alpha} =  
- c^{\alpha 3\gamma 3}w_{\gamma , 3 }  +  v^{\alpha 3\gamma 3 3 3}
w_{\gamma , 3 , 3 , 3 }
- c^{\alpha 3\gamma (\delta)}\hat{u}_{\gamma , (\delta)} + 
c^{\alpha 3\gamma\delta}\hat{u}_{\varepsilon}
\Gamma^{\varepsilon}_{\gamma\delta}
+ v^{\alpha 3\gamma\delta\varepsilon\zeta} \hat{u}_{\gamma ( , \delta , \zeta , \varepsilon)} 
+ v^{\alpha (\beta) \gamma 3 3 3}\hat{u}_{\gamma , 3 , 3 , (\beta)} 
- h^{\alpha\beta}\hat{u}_{\beta , 3 , 3}  \, .
\end{equation}
To compensate  the smallness of $ v^{\alpha \beta \gamma \delta \varepsilon \zeta} \, $  in the fifth term of this equation there should be  at least two differentiations in $x^3$. Thus after some reductions  one can rewrite (\ref{eq:sdefm5}) as  follows:
\begin{equation}
\label{eq:sdefm8}
s^{\alpha} =   
- c^{\alpha 3\gamma 3}w_{\gamma , 3 }  +  v^{\alpha 3\gamma 3 3 3}
w_{\gamma , 3 , 3 , 3 }
- c^{\alpha 3\gamma (\delta)}\hat{u}_{\gamma , (\delta)} + 
c^{\alpha 3\gamma\delta}\hat{u}_{\varepsilon}
\Gamma^{\varepsilon}_{\gamma\delta} 
+ 2( v^{\alpha 3\gamma (\beta) 3 3 } +  
v^{\alpha 3  \gamma 3 (\beta) 3})\hat{u}_{\gamma , 3 , 3 , (\beta)} 
- h^{\alpha\beta}\hat{u}_{\beta , 3 , 3}  \, .
\end{equation}

It is clear that in (\ref{eq:sdefm8}) one should  use only corrections $w_{\gamma}$ of order $v^{1/2}\,$. Only such corrections yield a contribution of the same order as the other terms. To find the corresponding contributions, we should keep in the exact equation 
\begin{equation}
\label{eq:hatuweq}
c^{\alpha\beta\gamma\delta}(\hat{u}_{\gamma}+
w_{\gamma})_{;\delta;\beta}  -
v^{\alpha\beta\gamma\delta\varepsilon\zeta}
(\hat{u}_{\gamma}+w_{\gamma})_{;\delta;\zeta;\varepsilon;\beta} =
v^{\alpha\beta\gamma\delta\varepsilon\zeta}\tilde{u}_{\gamma;\delta;\zeta;\varepsilon;\beta} 
   \,  
\end{equation}
only the terms of the order  $v^{-1/2}\,$. Indeed if $w_{\gamma} \sim v^{1/2}\,$ then the main terms $w_{\gamma , 3 , 3}$ and $v^{\alpha 3\gamma 3 3 3}w_{\gamma , 3 , 3 , 3 , 3}$ have just  such an order. Note that although there are terms $\hat{u}_{\gamma , 3 , 3}$ and $v^{\alpha 3\gamma 3 3 3}\hat{u}_{\gamma , 3 , 3 , 3 , 3}$ in the equation, which are of the order $\sim v^{-1}\,$, now they   cancel each other completely due to the fact that  now $\hat{u}_{\gamma}$ denotes the main approximation. Right hand side of (\ref{eq:hatuweq}) is $\sim v\,$, so  it should be omitted. The terms $c^{\alpha\beta\gamma\delta}w_{\gamma (;\delta;\beta)}\,$ and  $v^{\alpha\beta\gamma\delta\varepsilon\zeta}w_{\gamma (;\delta ;\zeta ;\varepsilon ;\beta)}\,$ should be omitted by the same reasons.  Thus (\ref{eq:hatuweq}) is reduced to
\begin{equation}
\label{eq:hatuwsimpl}
- c^{\alpha 3\gamma 3}w_{\gamma , 3 , 3}  +
v^{\alpha 3\gamma 3 3 3}w_{\gamma , 3 , 3 , 3 , 3} =
c^{\alpha\beta\gamma\delta}\hat{u}_{\gamma (;\delta ;\beta)}  -
v^{\alpha\beta\gamma\delta\varepsilon\zeta}
\hat{u}_{\gamma (;\delta ;\zeta ;\varepsilon ;\beta)}
   \,  .
\end{equation}

Equation (\ref{eq:hatuwsimpl}) requires further simplification: in the right-hand side we should keep only the terms  $\sim v^{-1/2}\,$. For this in the first term there should be one partial derivative in $x^3\,$, and in the second term there should be three derivatives in $x^3\,$. This is why we simplify second and fourth covariant derivatives as:
\begin{equation}
\hat{u}_{\gamma ; \delta ; \beta} \approx
\hat{u}_{\gamma , \delta , \beta} -
\hat{u}_{\gamma , \rho}\Gamma^{\rho}_{\delta\beta} -
\hat{u}_{\rho , \delta}\Gamma^{\rho}_{\gamma\beta} -
\hat{u}_{\rho , \beta}\Gamma^{\rho}_{\gamma\delta} \, ,
\end{equation}

\begin{equation}
\begin{array}{l}
\displaystyle
\hat{u}_{\gamma ; \delta ; \zeta ; \varepsilon ; \beta} \approx 
\hat{u}_{\gamma , \delta , \zeta , \varepsilon , \beta} -
\hat{u}_{\rho , \delta , \zeta , \varepsilon} \Gamma^{\rho}_{\gamma\beta} -
\hat{u}_{\gamma , \rho , \zeta , \varepsilon} \Gamma^{\rho}_{\delta\beta} 
- \hat{u}_{\gamma , \delta , \rho , \varepsilon} \Gamma^{\rho}_{\zeta\beta} -
\hat{u}_{\gamma , \delta , \zeta , \rho} \Gamma^{\rho}_{\varepsilon\beta} -
\hat{u}_{ \rho , \delta , \varepsilon , \beta }\Gamma^{\rho}_{\gamma\zeta} 
-  \hat{u}_{ \rho , \zeta , \varepsilon , \beta}\Gamma^{\rho}_{\gamma \delta} - \\ \\
\displaystyle
\hat{u}_{ \rho , \delta , \zeta , \beta}\Gamma^{\rho}_{\gamma\varepsilon}
-  \hat{u}_{\gamma , \rho , \varepsilon , \beta}\Gamma^{\rho}_{\delta\zeta}
-  \hat{u}_{\gamma , \rho , \zeta , \beta}\Gamma^{\rho}_{\delta\varepsilon}
-  \hat{u}_{\gamma , \delta ,\rho , \beta}\Gamma^{\rho}_{\zeta \varepsilon}
\, .
\end{array}
\end{equation}
Eventually the simplified equation    (\ref{eq:hatuwsimpl}) takes the form:
\begin{equation}
\label{eq:endweq}
\begin{array}{l}
\displaystyle
- c^{\alpha 3\gamma 3}w_{\gamma , 3 , 3 } +
 v^{\alpha 3\gamma 3 3 3}w_{\gamma , 3 , 3 , 3 , 3 } 
 =(c^{\alpha 3 \gamma (\beta)} + c^{\alpha (\beta) \gamma 3})
\hat{u}_{\gamma , (\beta) , 3} -
c^{\alpha\beta\gamma\delta}\Gamma^3_{\delta\beta}\hat{u}_{\gamma , 3} 
 - (c^{\alpha\beta\gamma 3} + c^{\alpha 3 \gamma\beta})
\Gamma^{\varepsilon}_{\gamma\beta}\hat{u}_{\varepsilon , 3} - \\ \\
\displaystyle
2(v^{\alpha 3\gamma (\beta) 3 3} +
v^{\alpha (\beta) \gamma 3 3 3})
\hat{u}_{\gamma  , (\beta), 3 , 3 , 3 } 
+ 2( v^{\alpha\beta\gamma 3 3 3}\Gamma^{\rho}_{\gamma\beta} +
v^{\alpha 3\gamma\delta 3 3}\Gamma^{\rho}_{\gamma \delta})
\hat{u}_{ \rho , 3 , 3 , 3} + \\ \\
\displaystyle
(4v^{\alpha\beta\gamma\delta 3 3 } \Gamma^{3}_{\delta\beta} +
v^{\alpha\beta\gamma 3 \delta 3} \Gamma^{3}_{\delta\beta} +
v^{\alpha 3\gamma\delta 3 \beta}
\Gamma^{3}_{\delta\beta })\hat{u}_{\gamma , 3 , 3 , 3}
\, .
\end{array}
\end{equation}

Taking into account that near the body surface  the Christoffel symbols and material tensors  can be considered as constants, the equation (\ref{eq:endweq}) has a very specific form:  all of  its terms are differentiated with respect to $x^3$  at least once.
Therefore, there exists a particular solution of this equation obeys an equation with one less  derivation:
\begin{equation}
\label{eq:endweqred}
\begin{array}{l}
\displaystyle
- c^{\alpha 3\gamma 3}w_{\gamma , 3  } +
 v^{\alpha 3\gamma 3 3 3}w_{\gamma , 3 , 3 , 3  } 
 =(c^{\alpha 3 \gamma (\beta)} + c^{\alpha (\beta) \gamma 3})
\hat{u}_{\gamma , (\beta) } -
c^{\alpha\beta\gamma\delta}\Gamma^3_{\delta\beta}\hat{u}_{\gamma } 
 - (c^{\alpha\beta\gamma 3} + c^{\alpha 3 \gamma\beta})
\Gamma^{\varepsilon}_{\gamma\beta}\hat{u}_{\varepsilon } - \\ \\
2(v^{\alpha 3\gamma (\beta) 3 3} +
v^{\alpha (\beta) \gamma 3 3 3})
\hat{u}_{\gamma  , (\beta), 3 , 3  } 
+ 2( v^{\alpha\beta\gamma 3 3 3}\Gamma^{\rho}_{\gamma\beta} +
v^{\alpha 3\gamma\delta 3 3}\Gamma^{\rho}_{\gamma \delta})
\hat{u}_{ \rho , 3 , 3 } + \\ \\
\displaystyle
(4v^{\alpha\beta\gamma\delta 3 3 } \Gamma^{3}_{\delta\beta} +
v^{\alpha\beta\gamma 3 \delta 3} \Gamma^{3}_{\delta\beta} +
v^{\alpha 3\gamma\delta 3 \beta}
\Gamma^{3}_{\delta\beta })\hat{u}_{\gamma , 3 , 3 }
\, .
\end{array}
\end{equation}

Note that in the left-hand side of (\ref{eq:endweqred}) there is  exactly the expression that we need in  (\ref{eq:sdefm8}). The general solution of the inhomogeneous differential equation is the sum of a particular solution of the inhomogeneous equation and the general solution of the homogeneous equation. But the remarkable fact is that any solution of the homogeneous equation decaying away from the surface, when substituted into $ - c^{\alpha 3\gamma 3}w_{\gamma , 3  } +  v^{\alpha 3\gamma 3 3 3}w_{\gamma , 3 , 3 , 3  }$ yields zero.  So it is quite enough to consider only a particular solution, and we can simply replace $- c^{\alpha 3\gamma 3}w_{\gamma , 3  } + 
 v^{\alpha 3\gamma 3 3 3}w_{\gamma , 3 , 3 , 3  }$ in  (\ref{eq:sdefm8}) by right-hand side of (\ref{eq:endweqred}).  Having made the replacement, we obtain $s^{\alpha}$ of the form:
\begin{equation}
\label{eq:snotready}
\begin{array}{l}
\displaystyle
s^{\alpha} =  - c^{\alpha 3\gamma (\delta)}\hat{u}_{\gamma , (\delta)} + 
c^{\alpha 3\gamma\delta}\hat{u}_{\varepsilon}
\Gamma^{\varepsilon}_{\gamma\delta} 
+(c^{\alpha 3 \gamma (\beta)} + c^{\alpha (\beta) \gamma 3})
\hat{u}_{\gamma , (\beta) } -
c^{\alpha\beta\gamma\delta}\Gamma^3_{\delta\beta}\hat{u}_{\gamma } 
 - (c^{\alpha\beta\gamma 3} + c^{\alpha 3 \gamma\beta})
\Gamma^{\varepsilon}_{\gamma\beta}\hat{u}_{\varepsilon } - \\ \\
\displaystyle
2(v^{\alpha 3\gamma (\beta) 3 3} +
v^{\alpha (\beta) \gamma 3 3 3})
\hat{u}_{\gamma  , (\beta), 3 , 3  } 
+ 2 v^{\alpha 3\gamma (\delta) 3 3 } \hat{u}_{\gamma  , 3 , 3 , (\delta) }
+ 2 v^{\alpha 3  \gamma 3 (\beta) 3}\hat{u}_{\gamma , 3 , 3 , (\beta)} 
- h^{\alpha\beta}\hat{u}_{\beta , 3 , 3}  \, .
\end{array}
\end{equation}
Elementary transformations lead (\ref{eq:snotready}) to a rather simple form:
\begin{equation}
\label{eq:sready}
s^{\alpha} =  c^{\alpha (\beta) \gamma 3}\hat{u}_{\gamma , (\beta) } -
c^{\alpha\beta\gamma\delta}\Gamma^3_{\delta\beta}\hat{u}_{\gamma } 
 - c^{\alpha\beta\gamma 3} \Gamma^{\varepsilon}_{\gamma\beta}
 \hat{u}_{\varepsilon } - h^{\alpha\beta}\hat{u}_{\beta , 3 , 3} 
\, ,
\end{equation}
where $h^{\alpha\beta}$ is redefined as
\begin{equation}
\label{eq:hready}
h^{\alpha\beta} =  
v^{\alpha (\gamma) \beta 3 (\delta) 3 } \Gamma^3_{(\gamma)(\delta)} - 
 v^{\alpha 3\gamma 3 3 3}\Gamma^{\beta}_{\gamma 3} - 
2v^{\alpha 3 \beta\delta 3 3 } \Gamma^{3}_{\delta 3} 
- v^{\alpha\varepsilon\beta 3 \delta 3} \Gamma^{3}_{\delta\varepsilon}
 \, .
\end{equation}
This equations  together  with (\ref{eq:bond3m2})  completely  solves the problem under consideration.

\section{Conclusions}

In this paper we propose an approach to the description of  flexoelectric deformations of finite size bodies with  sufficiently smooth surfaces. This approach is based on the fact that in reality the higher elastic moduli are very small. This smallness allows us to represent the elastic displacement vector as a sum of classical and non-classical parts. For  the non-classical part  the explicit representation derived. As for the classical part, it can be found by solving the equations of the classical theory of elasticity with the boundary conditions  obtained in this work. These boundary conditions are also quite classical in the  form, so that  the classical part of the problem is quite standard and does not require a separate consideration.


\end{document}